# Ambient heat and human sleep

Kelton Minor[1,2,3]*, Andreas Bjerre-Nielsen[1,4], Sigríður Svala Jónasdóttir[5], Sune Lehmann[1,5], Nick Obradovich[6]*

[1]Copenhagen Center for Social Data Science, University of Copenhagen
[2]Department of the Built Environment, University of Aalborg
[3]Global Policy Laboratory, University of California, Berkeley
[4]Department of Economics, University of Copenhagen
[5]Department of Applied Mathematics and Computer Science, Technical University of Denmark
[6]Center for Humans and Machines, Max Planck Institute for Human Development

**Abstract:** Ambient temperatures are rising globally, with the greatest increases recorded at night. Concurrently, the prevalence of insufficient sleep is increasing in many populations, with substantial costs to human health and well-being. Even though nearly a third of the human lifespan is spent asleep, it remains unknown whether temperature and weather impact objective measures of sleep in real-world settings, globally. Here we link billions of sleep measurements from wearable devices comprising over 7 million nighttime sleep records across 68 countries to local daily meteorological data from 2015 to 2017. Rising nighttime temperatures shorten within-person sleep duration primarily through delayed onset, increasing the probability of insufficient sleep. The effect of temperature on sleep loss is substantially larger for residents from lower income countries and older adults, and females are affected more than are males. Nighttime temperature increases inflict the greatest sleep loss during summer and fall months, and we do not find evidence of short-term acclimatization. Coupling historical behavioral measurements with output from climate models, we project that climate change will further erode human sleep, producing substantial geographic inequalities. Our findings have significant implications for adaptation planning and illuminate a pathway through which rising temperatures may globally impact public health.

**Introduction**

Regular and sufficient sleep supports human physical and mental health[1]. Short sleep duration is associated with reduced cognitive performance[2], diminished productivity[3], increased absenteeism[4], compromised immune function[5], as well as elevated risk of hypertension, adverse cardiovascular outcomes, mortality[6], depression, anger and suicidal behaviors[7,8]. Acute sleep restriction delays reaction times[2], increases accident risk[9], inhibits the neural encoding of new experiences to memory[10], and limits the clearance of neurotoxic metabolites from the brain linked to aging and neurodegenerative diseases[11]. Nevertheless, growing proportions of industrialized populations do not obtain adequate sleep, a development attributed to lifestyle and environmental changes, but not yet fully understood[3,12,13]. Concurrently, nighttime ambient temperatures are increasing due to both anthropogenic climate change and the expansion of urban heat islands[14,15]. To inform policy, planning and design, more information is needed about the environmental factors that curtail or promote sufficient sleep, particularly the role played by outdoor ambient temperature[16,17].

Prior research investigating the influence of ambient temperature on sleep in adults has been largely restricted to short controlled laboratory studies or imprecise self-report surveys. Humans and other mammals have developed both neurophysiological and behavioral processes to coordinate rhythms of thermoregulation and sleep, presumably to conserve energy expenditure[18]. In humans, the maximal rate of core body cooling is strongly correlated with sleep onset, and sleep propensity peaks near the minimum of the core body temperature rhythm[19,20]. Preceding sleep onset, increased blood flow to the distal skin and extremities enables cooling of the core body temperature[21]. Both skin and core body temperatures become more sensitive to environmental temperature during sleep, and duration of wakefulness has been shown to increase when temperatures warm or cool outside of the thermoneutral zone – the range of ambient temperatures where the body can maintain its core temperature only through regulating dry heat loss (skin blood flow) – albeit under controlled conditions that constrain human adaptation[22].

Far less is known about the influence of outdoor ambient temperatures and meteorological conditions on adult sleep in real-world settings[17]. Evidence from self-report studies indicates that the prevalence of reported sleep deficiencies increases in warm weather[23–26]. The largest of these studies pooled data in the US from nationally representative health surveys and found that higher monthly nighttime temperature anomalies increased self-reported nights of insufficient sleep during the previous month. However, retrospective self-reported sleep outcomes are notoriously imprecise, unreliable, and have been shown to have questionable internal validity [27,28]. Thus, it remains an open question whether, and to what extent, ambient thermal and weather conditions might affect individual sleep duration and timing across a global adult population.

In contrast to the limited precision and resolution of measures employed by previous studies -- even the largest of which only used data from one country -- the global reach of commercial wearable devices holds promise for understanding the environmental determinants of human sleep. Here, we draw on a large-scale sleep dataset consisting of over seven million repeated sleep records spanning 68 countries from nearly 50,000 people using accelerometry-based fitness bands linked to a smartphone application (Fig. 1B, D). This sleep dataset replicates established age, interregional and socio-temporal sleep characteristics (Methods, Supplementary Tables 1-2, Fig. 1C). Accelerometry-based commercial wearable devices are increasingly ubiquitous and particularly well-suited for large-scale observational studies[29], offering several empirical advantages over previous research designs. In-situ sleep measures from fitness bands provide dynamic spatial and temporal reference information for precise merging with meteorological data across diverse geographic regions, enabling the study of the effect of temperature on within-individual changes across the entire sleep period. Moreover, objective measures of total sleep duration can be used to investigate if weather affects the probability of obtaining short sleep, following standard definitions[1].

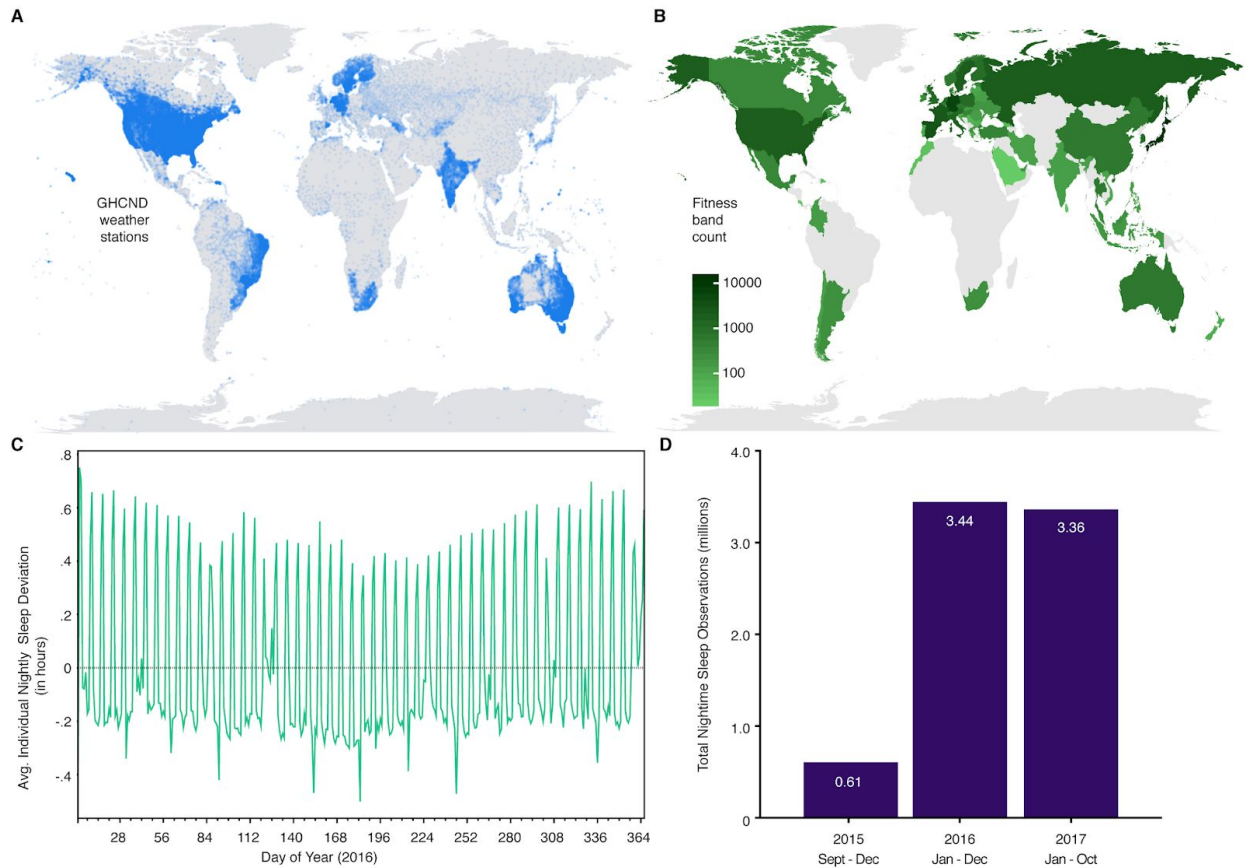

**Fig. 1** Global weather data coupled with over seven million daily sleep observations from fitness bands enables analyses of environmental factors that might influence sleep. **(A)**, Plotted map of weather stations from the Global Historical Climatology Network-Daily (GHCND). Each blue dot represents one station. **(B)**, World map depicting the country-level count of fitness band users included in this study, spanning 68 countries from all continents except for Antarctica. Countries with relatively more users appear as darker shades of green. **(C)**, Plot showing regular and dynamic temporal patterns in within-individual sleep duration deviation from average (in hours) over the 2016 calendar year. Each daily measure corresponds to the mean of all within-individual nightly sleep deviations for active users on that day. Recurring weekend peaks (above zero) and weekday valleys (below zero) reflect the imbalanced temporal structure of the adult working week - whereby sleep reduction during weekdays is partially compensated for on weekends with oversleep. **(D)**, Annual total number of nighttime sleep observations collected over the two-year period from September 2015 through October 2017, in millions.

To investigate if ambient temperature alters sleep behavior, we pair our fitness band observations of nighttime sleep duration (total sleep time) and timing (sleep onset and offset) with geolocated meteorological and climate data (Fig. 1A, B, Methods). We examine four questions. First, does ambient temperature impact human sleep duration and timing? Second, does temperature impact the probability of obtaining a short night of sleep? Third, do demographic and environmental factors moderate the susceptibility of human sleep to ambient temperature? Fourth, might climate change - through increased nighttime temperatures - affect sleep duration in the future?

To examine this first set of questions, we apply multivariate fixed-effects panel models with individual repeated measures, using exogenous variation in meteorological variables relative to local climate norms to estimate the acute effect of temperature and weather exposure on individual sleep outcomes (Methods, Supplementary Table 5). An advance of the present study is that our dataset allows us to control for all stable individual characteristics and leverage within-person fluctuations in both weather exposure and sleep outcomes to identify the causal effect of nighttime temperature on our person-level sleep outcomes of interest. Importantly, this fixed effects model controls for alterations in daylight hours, removing the potential confounding effect of daylight from our analyses (Supplementary Table 26). We verify that our primary conclusions are robust to alternative sample inclusion criteria, meteorological data, temporal controls and outcome measures (Methods, SI, Supplementary Tables 5-19, 26-27, Fig. 2). Further, our modelling framework controls for any unobserved, fixed device characteristics and we confirm that the period and frequency of wearable device use does not alter our primary results (Methods, Supplementary Tables 28-29).

**Results**

The results of our binned temperature regressions indicate that exogenous increases in nighttime ambient temperature reduce adult sleep duration across nearly the entire observed temperature range (Fig. 2A, Supplementary Fig. 2). Minimum temperatures exceeding 10°C are associated with a steeper decline in sleep duration. On average, nighttime temperatures registering above 25°C reduce individual sleep duration by over 7 minutes compared with 5 to 10°C (coefficient: -7.25, $p < 0.001$). Climate change is projected to continue to increase the magnitude and frequency of extreme nighttime temperatures beyond the recent historical record. Although data in our sample is sparse above 25°C, extending the range of temperature bins indicates that the warmest nights >30°C produce acute sleep loss of approximately 15 minutes per night compared to the coldest nights in our sample (Supplementary Fig. 3, Supplementary Tables 30, 31). Increasing nighttime temperatures amplify the estimated probability of obtaining a short night of sleep, measured with standard definitions for insufficient sleep[1]. The probability of sleeping less than seven hours increases gradually up to 10°C, before increasing at an elevated rate. Nighttime minimum temperatures greater than 25°C increase the probability of getting less than 7 hours of sleep by 3.5 percentage points compared with 5 to 10°C. Providing scale for this estimated relationship, a uniform shift from nighttime temperatures between 15°C and 20°C to 20°C and 25°C, if extrapolated for a population of 100,000 adults across a single night, would result in 1,100 additional individuals obtaining insufficient sleep. Conversely, colder ambient temperatures less than -5°C reduce the probability of insufficient sleep.

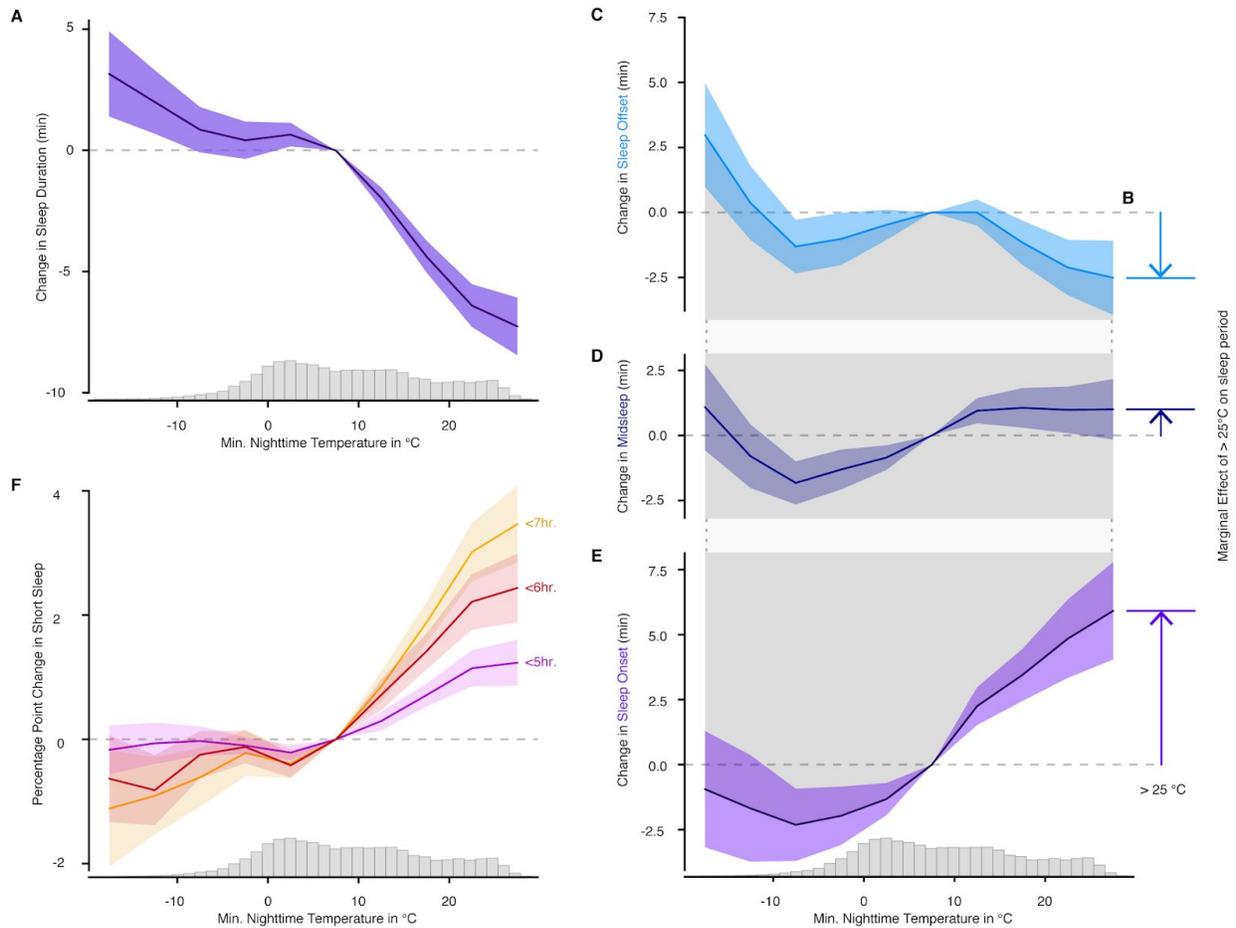

**Fig. 2** Increasing ambient temperature shortens the human sleep period, amplifying the probability of insufficient sleep attainment. **(A)** Plot of the relationship between increases in nighttime minimum temperature and the average within-individual change in sleep duration for each temperature bin. As minimum temperatures rise, sleep duration decreases with a steeper linear decline when temperatures exceed 10°C. Shaded regions represent 95% confidence intervals computed using heteroskedasticity-robust standard errors clustered on the first administrative division level. Histograms plot the distribution of observed nighttime temperatures across millions of sleep observations. **(B)** High nighttime temperatures >25°C significantly compress the human sleep period, primarily through a delay in sleep onset and a marginally smaller advance in sleep offset. **(C)** Sleep offset advances under higher nighttime temperatures >15°C, while very cold temperatures below -10°C delay offset timing. **(D)** Nighttime temperature increases above -10°C marginally delay midsleep - the midpoint of the human sleep period - although the magnitude of change at higher temperatures is smaller than concomitant changes in sleep onset and offset. **(E)** Above -10°C, increasing nighttime ambient temperatures delay sleep onset across the observed temperature distribution. **(F)** A plot of the predicted change in the probability of obtaining a short night of sleep across each minimum temperature bin. As temperature increases above 5°C, the probability of obtaining a short night of sleep - measured with three standard criteria - also increases.

Our results are robust, even when employing more extreme thresholds of insufficient sleep including <6hr and <5hr, demonstrating that marginal losses in total sleep time with rising temperatures predisposes people to insufficient sleep attainment (Fig. 2F, Supplementary Tables 8, 9). Further, since prior evidence from aggregated mobile phone calling data suggests that people may compensate for seasonal sleep reductions during the summer with afternoon naps, we check that our primary results are robust to replacing nighttime sleep duration with 24 hour sleep duration[30]. Contrary to the hypothesis that total sleep time might be conserved, we find that including daytime sleep actually slightly increases the effect size of temperature on within-individual sleep loss within our sample (Supplementary Tables 5, 25).

Our finding that human sleep is sensitive to increasing ambient temperatures across the temperature range differs from previous experimental studies that found reductions in sleep under both high and low environmental temperatures[22], though it extends findings from other observational, self-report studies[23,26]. In real-world settings, humans appear to be better at adapting their surroundings to obtain sufficient sleep under cooler conditions, whereas sleep loss increases with rising ambient temperatures. Since other meteorological factors may also influence sleep, we use our primary flexible model specification (Methods, Eq. 1b) to estimate the human sleep response to changes in weather. Sleep loss increases further as a function of the diurnal temperature range between maximum daytime and minimum nighttime temperature values (tmax - tmin). Since our specified model controls for other weather variables, including cloud cover and relative humidity, two plausible explanations are that indoor environments may retain heat gained during the day or that daytime heat may impart physiological demands that extend into the sleep period. By contrast, high levels of precipitation, wind speed and cloud cover each marginally increase sleep duration (Supplementary Fig. 1). Compared to moderate levels of relative humidity, both low and high levels reduce sleep, with the former producing greater sleep reduction.

To investigate how the entire sleep period responds to temperature-driven sleep loss, we construct separate flexible models to predict sleep onset, midsleep and offset timing. Drawing on these combined estimates, we illustrate that rising temperatures compress the human sleep period primarily through both a larger delay in sleep onset and a moderate advance in sleep offset. As minimum temperatures rise above -10°C, delays in sleep onset curtail sleep duration (Fig. 2A, C-E) and marginally delay midsleep. By contrast, nighttime temperature increases advance sleep offset timing when temperatures exceed 15°C. Nights with minimum temperatures greater than 25°C delay sleep onset by approximately six minutes (coefficient: 5.92, $p < 0.001$) and advance offset by about two and a half minutes (coefficient: -2.51, $p < 0.001$) compared to 5-10°C. Thus, larger declines in sleep duration at warmer nighttime temperatures are jointly driven by both delays in sleep onset and advances in sleep offset, constricting the human sleep period and slightly delaying midsleep (Supplementary Table 12).

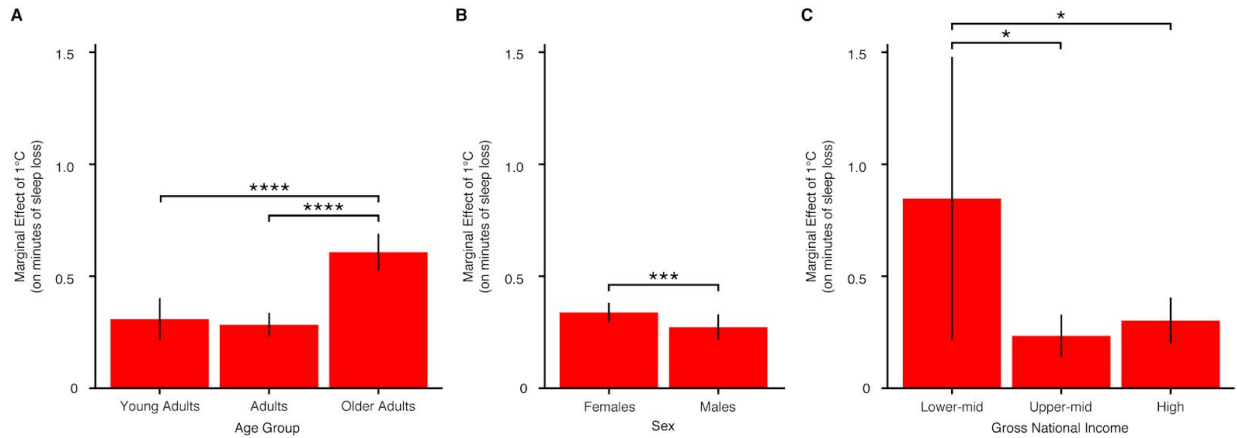

**Fig. 3** The effect of increasing ambient minimum temperature by 1°C on sleep loss is larger for older adults, females and people residing within lower-middle income countries. **(A)** The marginal effects of age category on sleep loss produced by interacting age group with nighttime minimum temperature within our primary model specification (Methods, Eq. 1d). The effect of increasing temperature by 1°C on sleep loss is nearly twice the magnitude for older adults (n = 155,922 observations), compared to mid-aged adults (n = 4,078,623 observations) and young adults (n = 170,626 observations). **(B)** The marginal effects of sex on temperature-driven sleep loss. Females (n = 1,279,271 observations) lose more sleep per degree increase in minimum temperature compared to males (n = 3,125,900 observations). **(C)** Plot of the marginal effect associated with interacting country-level gross national income (GNI) and nighttime minimum temperature. The effect of temperature on sleep loss is substantially larger for people residing within lower-middle income countries (n = 14,639 observations), compared to upper-middle (n = 274,488 observations) and high-income countries (n = 4,116,044 observations). Error bars represent 95% confidence intervals. Four stars indicate a significant difference at the p < .001 level, three stars indicate a significant difference at the p < .01 level, two stars indicate a significant difference at the p < .05 level and one star indicates a significant difference at the p < .1 level.

Individual and environmental demographic factors may modify the impact of temperature on sleep. Older adulthood is marked by an attenuated thermoregulatory response to suboptimal environmental temperatures, earlier sleep timing and reduced total sleep duration[31]. Such age-related developments may increase the nocturnal sensitivity of the elderly to higher ambient temperatures, possibly challenging sleep demand. We find that older adults (>65) are markedly more sensitive to exogenous increases in nighttime ambient temperature than mid-aged adults and young adults (Fig. 3A). The per degree effect of nighttime temperature on lost sleep for older adults (coefficient: -0.61) is over two times (p < .01) the effect estimated for mid-age adults (coefficient: -0.28). These results add to increasing evidence of the age-related ambient temperature sensitivity of sleep[23,32].

Under identical conditions, females' core body temperature decreases earlier in the evening compared to males[33], possibly exposing females to higher environmental temperatures around their time of habitual sleep onset. Females have also been shown to have greater subcutaneous fat thickness, which might impair nocturnal heat loss[24]. Comparing the effect of minimum temperature on sleep duration between sexes reveals that the per degree negative impact of nighttime temperature rise is larger (p < 0.01) for females (coefficient: -0.34) than males (coefficient: -0.27) in our dataset (Fig. 3b). This finding adds to evidence that females may be more predisposed to adverse heat effects on health than males[34,35].

Since access to infrastructure, cooling technologies and other unobserved environmental resources may plausibly modify the extent to which temperature impacts sleep, we further test if our results differ across country-income levels. We find that the effect of minimum nighttime temperature on human sleep loss is substantially larger for people residing within lower-middle income countries (coefficient: -0.85) compared to countries with higher income levels (Supplementary Tables 23a,b). The negative effect of nighttime temperature on sleep duration is 2.8 times greater (p = .087) for residents in lower-middle income countries compared to those from high income countries (coefficient: -0.30) and 3.6 times greater (p = .057) compared to upper-middle countries (coefficient: -0.23). Collectively, these results provide initial evidence that countries from all observed income levels are sensitive to the effect of ambient nighttime temperature on sleep, but the amount of sleep loss per degree increase may be disproportionately larger for people in lower-middle income countries.

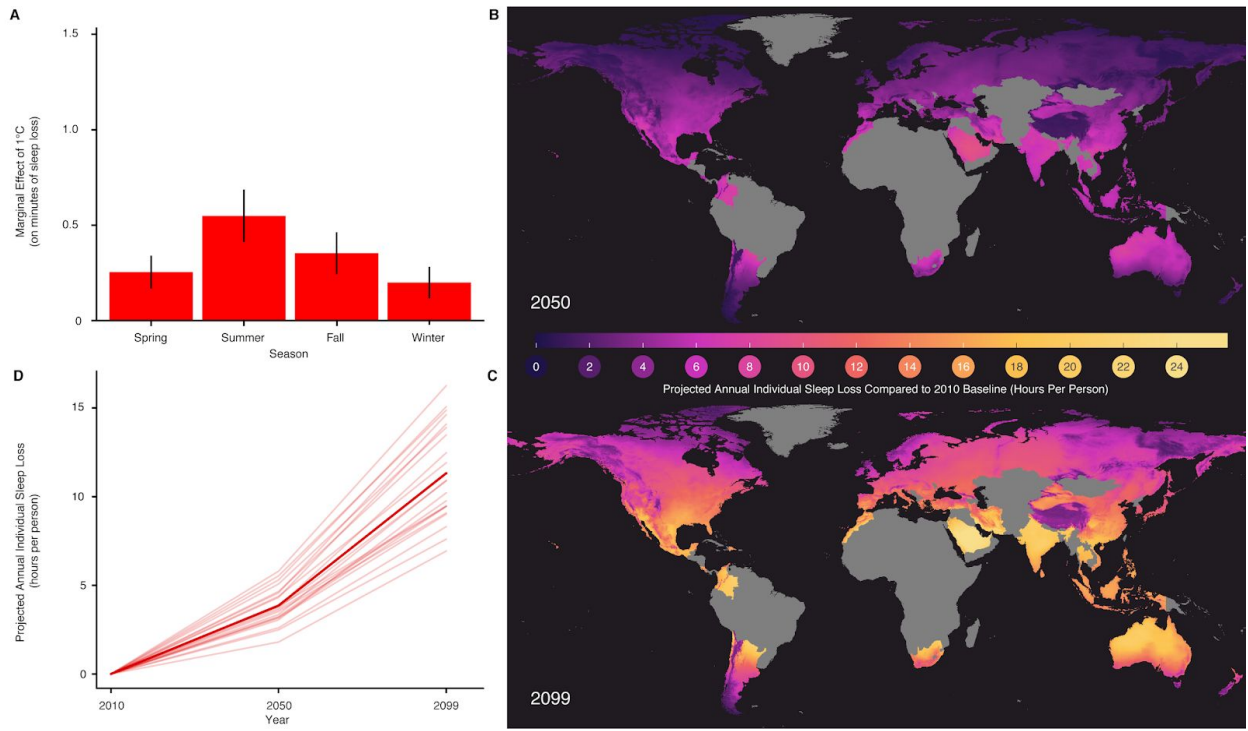

**Fig. 4** Climate change driven increases in nighttime temperatures may erode human sleep duration globally, with large seasonal and geographic variation. **(A)** The marginal effects of different seasons on nighttime minimum temperature-related sleep loss. Temperature increases are associated with the greatest sleep losses during summer nights, followed by fall, spring and winter nights. Error bars represent 95% confidence intervals. All marginal effects are significantly different from zero at the p < .01 level. **(B-C)** World maps projecting local annual sleep loss by 2050 and 2099 under a high emissions scenario. Each colored grid cell represents the additional per-person annual sleep loss projected for the corresponding 25 x 25km area. Darker purple colors represent areas with the lowest projected annual sleep losses, while lighter yellow colors signal areas with the largest annual sleep reductions. Geographic inequality in the magnitude of sleep loss is evident already by 2050 and becomes more pronounced by the end of the century. **(D)** Globally averaged individual-level projections for the impact of climate change on hours of sleep loss. Each line represents the estimated change in per-person sleep loss for a different downscaled climate model projection, averaged across all country-level pixels within the dataset. The dark red line plots the ensemble mean projected loss across 21 global climate models. Sleep loss increases over time due to projected warming across all countries (Supplementary Fig. 6).

To determine whether increases in minimum temperatures impact human sleep differently over the course of the year, we inspect the marginal effect of temperature on sleep loss across each season. Nighttime temperature increases result in sleep loss throughout the year (Fig. 4A). Consistent with the annual temperature distribution, we find that rising nighttime temperatures decrease sleep duration the most during summer months (coefficient: -0.55), followed by fall (coefficient: -0.35), spring (coefficient: -0.25) and winter (coefficient: -0.20) months (all coefficients are significantly different from zero at the $p < .01$ level). The per degree effect of an increase in nighttime temperature on sleep loss during the summer is nearly three times larger than in the winter. Our results provide further evidence that temperature increases impart the largest losses on human sleep when nighttime temperatures are already elevated (Fig. 2A, Supplementary Table 20). Since prior research suggests that people may be able to physiologically acclimatize to warmer temperatures over relatively short periods of time, we further test whether human sleep responds differently to nighttime temperature increases experienced during the first month of summer – when nights with locally hotter temperatures are relatively newer – versus the last month of summer when they are more familiar [36]. While short-run acclimatization would be apparent if the effect of temperature on sleep duration diminishes from the first to the last month of summer, we instead find evidence that nighttime temperatures appear to incur similar-to-marginally-more sleep loss near the end of summer when warmer temperatures are relatively less novel (Methods, Supplementary Table 35).

Finally, abiding by the assumption that future sleep loss will respond to expected changes in nighttime minimum temperature as they have responded to them in the recent past, we construct behavioral projections for the impact of climate change on cumulative annual sleep loss for all countries contained within our dataset (Fig. 1B). Consistent with the climate impacts literature, we draw upon gridded data from 21 global climate models run under a high emissions scenario (RCP8.5) and link downscaled, nighttime temperature projections to compute the per-person average annual sleep loss expected by the middle and end of the century (Fig. 4B-D, Methods). Averaging across all grid cells and each of the 21 climate models, we find that annual individual sleep loss due to warming nighttime temperatures may increase by mid-century, with yearly losses becoming larger by 2099. By the end of the century, residents from a majority of the earth's landmasses may lose over 11.3 hours of sleep per year due to climate change, amounting to over one and a half average night(s) of lost sleep (Methods). Due to variation in climate model output, global average sleep loss projections for 2099 range from a minimum of 6.93 hours to a maximum projected annual individual loss of 16.25 hours. Since projected changes in nighttime temperatures also vary geographically, we map each of the projected grid cells for all countries in our dataset (Fig. 4B). Without further adaptation, residents in the warmest areas are projected to experience cumulative sleep losses of over 23 hours per year by 2099, equivalent to over three average nights of sleep for our sample (Fig. 4C, Supplementary Fig. 6, Methods). Importantly, our historical estimate underlying these projections may be conservative since the majority of data arise from high income countries and are skewed towards a middle-aged, male demographic (Methods). Indeed, our marginal effect estimates indicate that future sleep loss may be considerably larger for lower income countries, demographically older populations and women (Fig. 3).

**Discussion**

In summary, we provide extensive evidence that human sleep is sensitive to nighttime ambient temperature, posing an additional climate change related threat to global public health. Increases in nighttime minimum temperature reduce sleep duration and increase the probability of obtaining insufficient sleep via the constriction of the human sleep period, primarily through a delay in sleep onset. The effect of nighttime temperature on sleep loss is amplified for lower income countries, older adults and females. Our results suggest that temperature-driven sleep losses are evident across demographics, and increasing temperatures lead to some within-person sleep loss across all seasons, with the largest losses during the warmest months and on nights when minimum temperatures exceed 10°C. We do not find evidence of possible short-term acclimatization of sleep to warmer temperatures. Further, increasing nighttime temperatures may erode human sleep into the future. The burden of future warming will not be evenly distributed barring further adaptation, with people living in warmer climates expected to lose considerably more nights of sleep per year by 2099. Taken together, our results demonstrate that temperature-driven sleep loss likely has and may continue to contribute to global inequalities in environmental health.

Our results carry significant implications for adaptation planning, policy and research. Growing evidence faults increases in temperature with societal impacts to public health and behavior, although the causal mechanisms have remained poorly characterized[37–42]. Insufficient sleep increases the risk of many of the same negative behavioral, social and economic outcomes shown to increase with high temperatures[35,43–48]. Thus, sleep may act as a key mediator between ambient temperature and adverse cognitive and behavioral outcomes[2,4–9,26]. While further research should seek to clarify this hypothesis, addressing the nocturnal impact of rising ambient temperatures on human sleep may be an efficient early intervention to reduce downstream adverse behavioral impacts linked to insufficient sleep. Through the use of consistently measured sleep records registered by wearable devices, our findings indicate that elevated temperatures drive sleep loss primarily by delaying when people fall asleep, providing a specific behavioral target for future adaptive interventions that seek to attenuate the impact of nighttime heat.

Interestingly, a corollary to our results is that ambient cooling interventions may be able to promote sleep gain (Fig. 2A). Although access to air conditioning may partially buffer the effect of high ambient temperatures (Fig. 3C), these same adaptive technologies can potentially exacerbate the unequal burdens of both global and local warming, through increased greenhouse gas emissions and ambient heat displacement[15,39,49,50]. Societal innovations in thermal management, planning and design may be needed to protect the world's urban population centers and vulnerable communities from exposure to magnified nighttime temperatures[42].

Several considerations should be taken into account when interpreting our results. First, global access and adoption of wearable devices is not geographically or demographically uniform. Our dataset contains more people who are middle-aged, male and from high and upper middle-income countries (Fig. 1B, Methods). Given that nighttime temperature effects are larger for females, the elderly and lower-middle income countries in our sample, the magnitude of our effect estimates and projections are likely conservative. Wearable device ownership may also be associated with unobserved demographic factors including higher socioeconomic status and access to cooling technologies, possibly reducing the accuracy of our estimates — especially in lower-middle income countries[50]. Further, people who select into buying fitness bands may be more physiologically resilient.

Second, station-based measures of ambient temperature may differ from actual temperature exposures where people live, likely attenuating the magnitude of our empirical estimates of the relationship between temperature and sleep[20,25,51]. As such, our results reflect the total effect of ambient outdoor temperature on human sleep duration and timing, including all sleep-adjacent behavioral effects. Moreover, the deconvolution process used in our analyses likely further downward biases our effect estimates (Methods). Nevertheless, we observe consistent ambient temperature effects on sleep - even for people living in industrialized societies and high-income countries with plausible access to air conditioning (Fig. 3C).

Third, the current study exclusively measured changes in sleep duration and timing and thus does not convey how the observed decline in sleep duration impacts underlying sleep physiology. Controlled experiments with human subjects have shown that REM and NREM sleep decrease when people are exposed to high environmental temperatures[22], however it remains unclear how ambient temperature modulates human sleep architecture and other neurobehavioral correlates of restorative sleep in real-world settings[19]. Further, similar to wrist actigraphy, accelerometer-based wearable devices may underestimate nighttime awakenings, suggesting that the sleep estimates in this study may be conservative. Future in-situ research should investigate whether sleep fragmentation is also sensitive to ambient weather conditions.

Fourth, although our sample includes data from 68 countries spanning all populated continents, it has sparse coverage for large parts of Africa, Central America, South America and the Middle East - regions that already rank among the warmest in the world (Fig. 4C). Climate projections indicate that many of the countries within these regions will be disproportionately exposed to some of the highest ambient temperatures and most cooling degree days, warranting future study[49].

Lastly, even though we do not find evidence of short-term acclimatization, it is possible that people may adapt to warmer nighttime temperatures in the future through technological or environmental developments not captured by our historical estimates, which likely already reflect considerable adaptation (Fig. 3C) [52,53]. To this end, future research is needed to investigate equitable policy, planning and design innovations that alleviate the stress of elevated nighttime temperatures and promote resilient slumber on both a local and societal scale.

**Materials and Methods**

No statistical methods were used to predetermine sample size. Datastreams from wearables and mobile devices hold potential utility for assessing the human impacts of global environmental changes[54,55]. For our user-level analysis, we include sleep entries collected over a two-year period from September 2015 through October 2017 from 47,628 anonymous fitness band users who electronically consented for their data to be processed for research purposes. To our knowledge, this is the largest sample of mobile sleep tracking device users yet employed to study the relationship between meteorological factors and human sleep. To investigate if meteorological variables influence several sleep parameters of interest, we register the following for each user in our dataset: the onset time when the sleep period commences (sleep onset); the midpoint of the registered sleep period (midsleep); the detected time when the sleep period ends (sleep offset); and the total sleep time registered during a given night (sleep duration).

During a preprocessing stage, self-reported age, sex, height and weight data were aggregated into age group, sex and WHO BMI categories and sleep entries were matched to the nearest weather stations. 10.67 billion sleep state observations - measured in 1 minute epochs - were aggregated into 7.41 million sleep records. We adopt inclusion criteria for min and max allowable sleep duration used in prior global observational sleep studies (4 hrs. < sleep duration < 12 hrs.)[56]. To study the effect of nighttime ambient temperature on concurrent sleep attainment during the nocturnal period, we further filter sleep entries based on local timing (19:00 < sleep onset time < 08:00 and 00:00 < sleep offset time < 15:00). Our primary results are robust to using alternative sleep timing filters used in the literature instead of these wider criteria (Supplementary Table 27)[56,57]. Since people may compensate for insufficient sleep at night with daytime naps shorter than four hours, we confirm that our primary results are robust to using 24-hour sleep attainment instead (Supplementary Table 5).

To ensure sufficient temporal coverage, we require each person to have a minimum of four weeks (28 nights) with sleep entries and registered sleep observations on at least 25% of nights spanning the period from first to last use. Our results are robust to alternative temporal inclusion criteria, including constraining our analysis to only people with a minimum of 56, 84 and 112 total nights with sleep entries (Supplementary Table 29). Furthermore, our results persist when constraining our sample to those with regular wearable device use on 50%, 75% as well as 85% of nights from their date of first use (Supplementary Table 28). The data was registered by waterproof fitness-tracking wristbands that utilize an internal accelerometer to detect movement and measure sleep and wake states at the minute-level. The wristbands were internally validated at temperatures exceeding those observed in the present study and have a listed operating temperature range of -20°C to 60°C, well above the distribution of ambient temperatures encountered in this study. The sleep and wake state estimates produced by these fitness bands have been found to accurately align with independent, contemporaneous measurements of mobile device inactivity and activity, a behavioral proxy for wakefulness [58]. We further assess the external validity of sleep estimates produced by these devices by comparing the national and demographic sleep estimates generated by the fitness bands to results from a suite of previously published sleep studies [59]. The resulting global dataset reproduces established socio-temporal, demographic and geographic sleep trends. Distinct weekend-weekday differences in sleep duration are indicative of characteristic work and social schedules that constrain human sleep, with reduced sleep on weekdays and sleep recovery on weekends (Fig. 1C, Supplementary Table 2). For this sample of wearable device users, the median individual average nighttime sleep duration was 7.1 hours. A greater percentage of older adults (43.6%) regularly slept less than 7 hours a night compared to young adults (32.7%), and middle-aged adults had shorter average sleep duration during the working week compared to both younger and older adults, consistent with the literature[60] (Supplementary Tables 1 – 2). Previous research also suggests that average sleep duration in Eastern countries is moderately lower than in Western countries[56,61,62]. Our wearable dataset replicates this result, with adults in Japan sleeping less on both weekdays and weekends compared to adults from four different European countries, across all adult age ranges (Supplementary Table 2).

Our sample consists of people who self-selected into wearable device use and thus may differ from the background population in particular ways. We observe that our sample consists of a greater proportion of males (~69%) than females (~31%). Further, participants in the dataset resided entirely within high and middle-income countries. ~80% of included users were from 42 different high-income countries and ~20% were from 26 middle income countries, 9 of which were lower middle-income countries. The sample consists primarily of middle-aged adults (25 - 65 yrs., ~91%), with fewer young-adults (19-25 yrs., 6%) and older adults (65+ yrs., 3%). However, the median age values from the top five countries with the most users in our dataset closely correspond to the UN standard population median values for the same countries, with an average absolute difference of 2.6 years (Supplementary Table 3). Furthermore, the age-standardized BMI values from the top five countries with the most users in our dataset fall within or near the WHO population estimate ranges for these countries (Supplementary Table 4). Of note, the age-standardized BMI for both men and women from Japan was slightly higher than the WHO range. We do not find evidence for heterogeneity in the effect of nighttime temperature on sleep duration across BMI categories in our dataset (SI Figure 5).

**Multivariate Fixed Effects Panel Models**

To investigate whether ambient temperature and outside weather alter human sleeping behavior, we link individual nightly sleep observations with meteorological data from two sources. For each nighttime sleep observation, we utilize the proximity weighted average of nearest station-level minimum nighttime temperature and precipitation data within a 100km radius from the National Centers for Environmental Information Global Historical Climatology Network - Daily (GHCND) output as well as wind speed, daily cloud cover and relative humidity data from the NCEP Reanalysis 2 project[63,64].

Our primary relationship of interest is the effect of daily nighttime minimum temperature on within-person changes in sleep duration.

$$Y_{ijktm} = TMIN_{ijktm} + TMINORM1981to2010_{ijktm} + Z\eta + \alpha_i + \mu_t + \nu_{km} + \varepsilon_{ijktm} \quad (1a)$$

In this multivariate fixed effects linear model, $i$ indexes individuals, $j$ indexes 2nd-level administrative division, $k$ indexes 1st-level administrative division, $t$ indexes unique calendar date and $m$ indexes unique calendar months. Our dependent variable $Y_{ijktm}$ represents the sleep duration (in minutes) of individual $i$, in a given 2nd admin division $j$, 1st admin division $k$, on calendar day $t$ and calendar month $m$. Our independent variable of interest is minimum nighttime temperature, $TMIN_{ijktm}$. We also control for the local historical minimum temperature climate normal from 1981 to 2010, $TMINORM1981to2010_{ijktm}$. Further, we control for daily precipitation, daily temperature range ($TMAX_{ijktm}$ - $TMIN_{ijktm}$), percentage cloud cover, relative humidity, and average wind speed, represented via $Z\eta$, as failure to do so may bias our estimates of the effect of nighttime minimum temperature on our sleep outcome measure[65]. Finally, we also include climate normals within $Z\eta$ for each of our meteorological control variables, computed as the local historical average value from 1981 to 2010 for a given user-date using the nearest weather stations.

Unobservable user-specific, geographic, or temporal factors may influence sleep outcomes. To ensure that these user-specific factors do not bias our estimates of the effect of weather on sleep duration, we include $\alpha_i$ - representing user-level indicator variables - in *Eq. 1a*. These variables control for all stable unobserved characteristics for each user and wearable device[65]. Further, there may be unobserved daily developments or region-specific seasonal changes - such as daylight - or secular trends influencing sleeping outcomes that might spuriously correlate with the weather. In order to control for these potential confounders, we include $\mu_t$ and $\nu_{km}$ in *Eq. 1a*, representing calendar date and 1st admin-by-calendar month indicator variables, respectively. Our primary results are robust to alternative temporal controls, including replacing 1st admin-by-month with 1st admin-by-week indicator variables to control for administrative region-specific weekly changes in daylight (Supplementary Table 26).

Our empirical identifying assumption, consistent with the climate econometrics literature[65,66], is that the remaining variation in daily minimum temperature is as good as random after conditioning on these fixed effects[67]. The estimated model coefficients from $f(TMIN_{ijktm})$ can thus be interpreted as the causal effect of minimum nighttime temperature on sleep duration[40,68].

We estimate *Eq. 1a* using ordinary least squares and adjust for possible spatial and serial correlation in $\varepsilon_{ijktm}$ by employing heteroskedasticity-robust standard errors clustered at the 1st administrative division level. We omit non-climatic control variables from *Eq. 1a* because of their potential to generate bias in our parameters of interest[65,69,70].

The global weather data that we rely on for our primary historical estimate has a higher concentration of stations providing temperature and precipitation measurements over North America and Eurasia than over Africa and parts of South America, resulting in some exclusion of users into the final sample for our primary model and potentially less precise ambient temperature estimates for lower-middle income countries. Our results are robust to using globally gridded reanalysis data instead (SI Fig. 2A; Supplementary Tables 5, 6)[63]. Importantly, the resulting empirical estimates from our analyses are likely conservative. A combination of measurement error and amplification of noise due to the deconvolution process we use in our analyses likely further downward biases our estimates of the relationship between temperature and sleep behavior[51,71,72].

$$Y_{ijktm} = f(TMIN_{ijktm}) + TMINORM1981to2010_{ijktm} + Z\eta + \alpha_i + \mu_t + \nu_{km} + \varepsilon_{ijktm} \quad (1b)$$

In *Eq. 1b*, we convert minimum nighttime temperature in *Eq. 1a* into indicator variables for each 5°C minimum temperature and temperature range bin[73], 1cm precipitation bin, 5 m/s windspeed bin, as well as each 20 percentage point bin of cloud cover and relative humidity. This enables us to flexibly estimate a non-linear relationship between each of our meteorological variables and sleep outcomes[36] (Fig 2A., SI Fig. 1). We omit the 5°C to 10°C minimum temperature, 5°C to 10°C daily temperature range, 0 cm precipitation, 0-5 m/s windspeed, 0% cloud cover and 60-80% humidity bins as reference categories. We interpret our estimates as the average within-individual change in sleep duration at a particular temperature bin relative to these baselines. Our results are consistent when binning each of our climate control variables as well (Supplementary Table 32-34). Our flexible model results are robust to substituting the National Weather Service (NWS) Heat Index[74] -- a measure of heat stress -- for minimum temperature and relative humidity Eq. 1b. (SI Fig. 2B, Supplementary Table 15). The flexibly estimated functional form produced by our primary specification described by *Eq. 1b* persists when extending the range of temperature bins to model the effects of extremely warm and cold nights, despite sparse data coverage in these binned extremes (SI Fig. 3, Supplementary Tables 30, 31).

$$Y_{ijktms} = TMIN_{ijktm} * \varnothing_s + TMINORM1981to2010_{ijktm} + Z\eta + \alpha_i + \mu_t + \nu_{km} + \varepsilon_{ijktm} \quad (1c)$$

In order to inspect the marginal effects of season of the year on our relationship of interest, we preserve the specification of *Eq.1a*, and interact our measure for nighttime minimum temperature $TMIN_{ijktm}$ with a categorical variable representing season $\varnothing_s$. We interpret the resulting estimates as the marginal effects of a 1°C minimum temperature increase on sleep loss, for a particular season of the year (Supplementary Table 20).

$$Y_{ijktms} = TMIN_{ijktm} * \gamma_i + TMINORM1981to2010_{ijktm} + Z\eta + \alpha_i + \mu_t + \nu_{km} + \varepsilon_{ijktm} \quad (1d)$$

In order to inspect possible heterogeneity in the estimated effect of minimum temperature on sleep across demographics (ex. Young Adults vs. Adults vs. Older Adults), we add a demographic interaction term $\gamma_i$ to *Eq.1a* while maintaining the same model specification. We construct separate models for each interaction term (Fig. 3, SI Fig. 5, Supplementary Tables 21 - 24). Since some users did not self-report their age, sex or BMI information, our sample size in these regressions varies.

$$Y_{ijktms} = TMIN_{ijktm} * S_m + TMINORM1981to2010_{ijktm} + Z\eta + \alpha_i + \mu_t + \nu_{km} + \varepsilon_{ijktm} \quad (1e)$$

To test for possible short-term acclimatization to warmer nighttime temperatures, we extract a subset of the data consisting of the first and last summer months for each location and year of observation. Thus, for observations originating from the northern hemisphere (southern hemisphere), June (December) is labelled as the first month of summer when locally warmer temperatures are a relatively newer occurrence while August (February) is labelled as the last month of summer when elevated temperatures have become more familiar. We add a summer month interaction term $S_m$ to *Eq.1a* while otherwise keeping the same model specification. The resulting estimates represent the marginal effects of a 1°C minimum temperature increase on sleep loss, for a particular summer month (Supplementary Table 35).

$$Q_{ijktm} = f(TMIN_{ijktm}) + TMINORM1981to2010_{ijktm} + Z\eta + \alpha_i + \mu_t + \nu_{km} + \varepsilon_{ijktm} \quad (2a)$$

For *Eq. 2a*, we replace sleep duration in *Eq. 1b* with sleep onset $Q_{ijktm}$ as our dependent variable, while otherwise maintaining the same flexible model specification (Supplementary Tables 12, 13).

$$R_{ijktm} = f(TMIN_{ijktm}) + TMINORM1981to2010_{ijktm} + Z\eta + \alpha_i + \mu_t + \nu_{km} + \varepsilon_{ijktm} \quad (2b)$$

For *Eq. 2b*, we instead specify midsleep $R_{ijktm}$ as our dependent variable, while maintaining the same flexible model specification as *Eq. 1b* (Supplementary Tables 12, 13).

$$U_{ijktm} = f(TMIN_{ijktm}) + TMINORM1981to2010_{ijktm} + Z\eta + \alpha_i + \mu_t + \nu_{km} + \varepsilon_{ijktm} \quad (2c)$$

For *Eq. 2c*, we use sleep offset $U_{ijktm}$ as our dependent variable, while preserving the same flexible model specification as *Eq. 1b* (Supplementary Tables 12, 13).

$$V_{ijktm} = f(TMIN_{ijktm}) + TMINORM1981to2010_{ijktm} + Z\eta + \alpha_i + \mu_t + \nu_{km} + \varepsilon_{ijktm} \quad (3)$$

In this multivariate fixed effects linear probability model, we preserve our flexible model specification in *Eq. 1b* but instead utilize a binary outcome $V_{ijktm}$ representing whether an individual (i), in county (j), in 1st-admin (k) on calendar day (t) within calendar month (m) attained a short night of sleep less than a standard threshold for insufficient sleep[75]. We interpret this estimate as the change in probability of obtaining insufficient sleep (Fig. 2F) relative to the meteorological baselines outlined in Eq. 1b. Our results are robust to employing alternative measures of short sleep (Supplementary Tables 8, 9).

$$Y_{ijktm} = TMIN.ANOMALY1981to2010_{ijktm} + Z\eta + \alpha_i + \mu_t + \nu_{km} + \varepsilon_{ijktm} \quad (4a)$$

As an alternative specification to *Eq. 1a*, we replace both $TMIN_{ijktm}$ and $TMINORM1981to2010_{ijktm}$ with a single term: nighttime temperature anomalies $TMIN.ANOMALY1981to2010_{ijktm}$ — computed as the difference between the two previous terms — while preserving the same multivariate fixed effects linear model specification as *Eq. 1a*. This new independent variable of interest represents the nightly minimum temperature deviation from the normal historical average (from 1981 to 2010) for each user-night (Supplementary Table 17).

$$Y_{ijktm} = f(TMIN.ANOMALY1981to2010_{ijktm}) + Z\eta + \alpha_i + \mu_t + \nu_{km} + \varepsilon_{ijktm} \quad (4b)$$

For *Eq. 4b*, we adopt the same anomaly specification as *Eq. 1c* but include 1°C flexible temperature bins to nonparametrically estimate the relationship between nighttime minimum temperature anomalies and sleep duration. Furthermore, we add binned meteorological controls, employing the same baseline categories as specified in *Eq. 1b*. We omit the temperature anomaly reference range of -.5°C to .5°C and interpret our estimates as the average within-individual change in sleep duration within a particular nighttime temperature anomaly range relative to the binned meteorological baselines (SI Fig. 4, Supplementary Table 18).

**Spline Regression Model**
To investigate how climate change may impact human sleep in 2050 and 2099, we draw upon data from 21 Coupled Model Intercomparison Project Phase 5 (CMIP5) models[76] run on the Representative Concentration Pathway "high emissions" scenario RCP8.5[77] and extract NASA Earth Exchange Global Daily Downscaled Projections (NEX-GDDP) bias-corrected, statistically downscaled, nightly temperature projections[78] for each 25km x 25km grid cell spanning each country included in our global sleep dataset. Climate change is projected to extend the nighttime temperature distribution rightwards, resulting in extreme ambient temperatures that exceed our historical observations. Rather than assigning these extreme temperatures the fitted value from the highest temperature bin within the historical distribution, we fit a linear spline to the data — with knots placed at -20°C and 10°C — and forecast behavioral estimates for projected temperatures for 2050 and 2099[73]. The functional form yielded by the spline model closely mirrors the relationship uncovered by Eq. 1b.

**Person-Level Annual Projection Plots**
To plot the projected average individual change in sleep duration due to climate change (Fig. 4D), we extract the 2010, 2050 and 2099 average daily minimum temperature predicted sleep loss from all 21 of NASA's NEX-GDDP bias-corrected, statistically downscaled daily climate models. For each grid-cell-day-of-the-year and model, we subtract the fitted values in 2050 and 2099 from the reference period of 2010. This operation results in an estimate of per-person change in sleep loss for each grid-cell-day-of-the-year, for each model. We then sum these daily difference values over the year for each grid-cell, yielding a cumulative estimate of the projected annual sleep loss for each grid-cell and model combination. We then average across all global grid cells - correcting for geographic distortion - to compute the equal-area weighted average annual sleep loss for each model (Fig. 4D). Separately, we compute and plot country-level projected annual sleep loss (per-person) by instead averaging grid cells for each country, across all 21 downscaled climate models (SI, Fig. 6).

**Grid Cell Annual Projection Map**
To map the global grid-cell projections, we plot the ensemble mean estimate of annual per-person sleep loss across all models, for each grid cell (Fig. 4B, C). Thus, each colored grid cell represents the projected annual individual sleep loss for a person residing within the area demarcated by that grid cell, relative to the baseline period of 2010.

**Supplementary Materials:**

Figures S1-S6

Tables S1-S35

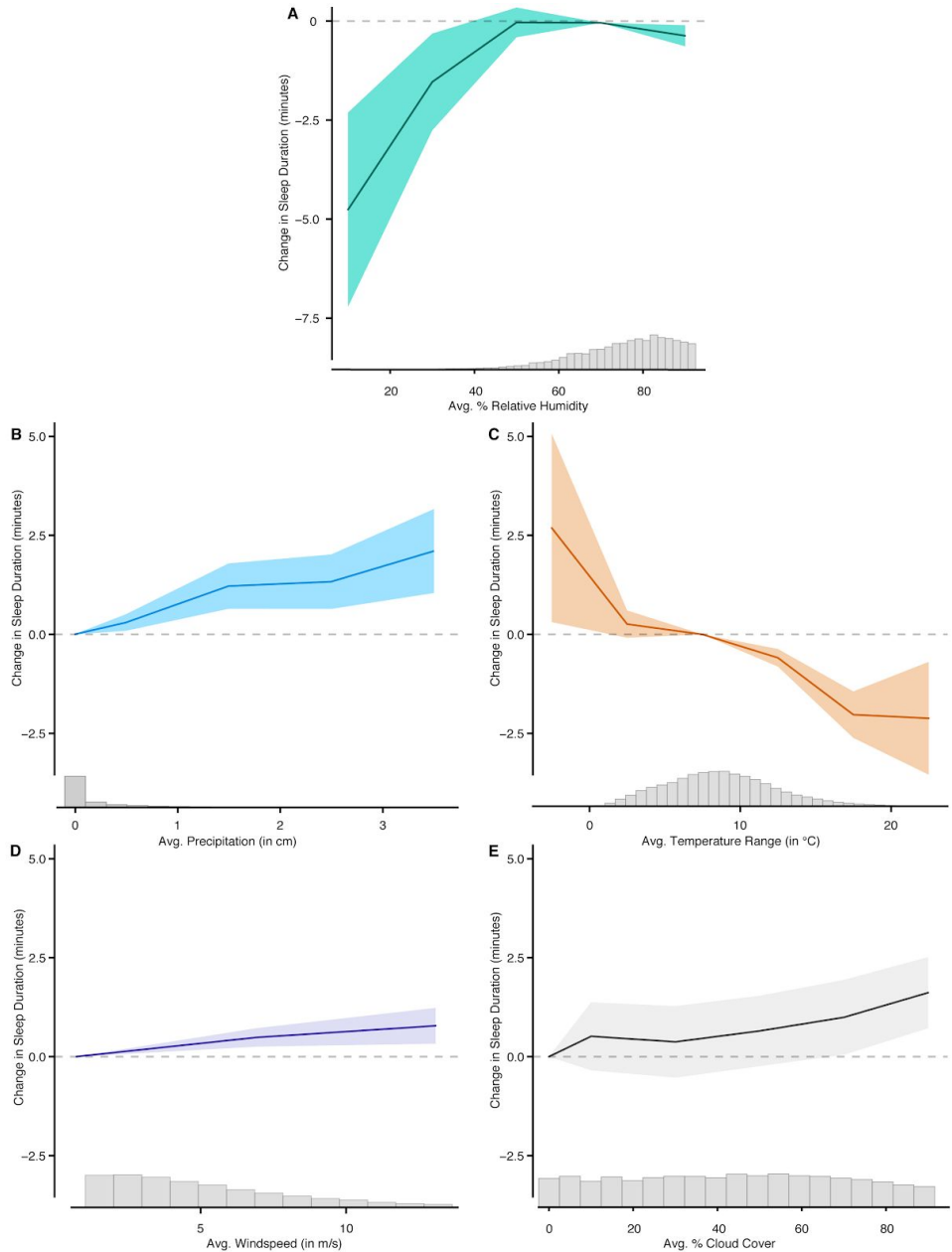

**SI Figure 1** Meteorological conditions influence sleep duration. **(A)**, Depicts the predicted within-individual change in sleep duration as a function of relative humidity. Low levels of relative humidity reduce sleep more than high relative humidity. Shaded regions represent 95% confidence intervals computed using heteroskedasticity-robust standard errors clustered on the first administrative division level. Histograms plot the distribution of observed meteorological variables spanning over 7 million sleep observations. **(B)**, Sleep duration increases as a function of precipitation intensity. **(C)**, An increase in the temperature range between the daily maximum and minimum temperature reduces sleep duration. **(D)**, Moderate to high wind speeds extend sleep duration. **(E)**, High cloud cover marginally increases sleep duration. The magnitude of the effect of moving across the distributions of these meteorological control variables is comparatively smaller than moving across the distribution of nighttime minimum temperature (Fig. 2A, SI Fig. 2A, Supplementary Table 6).

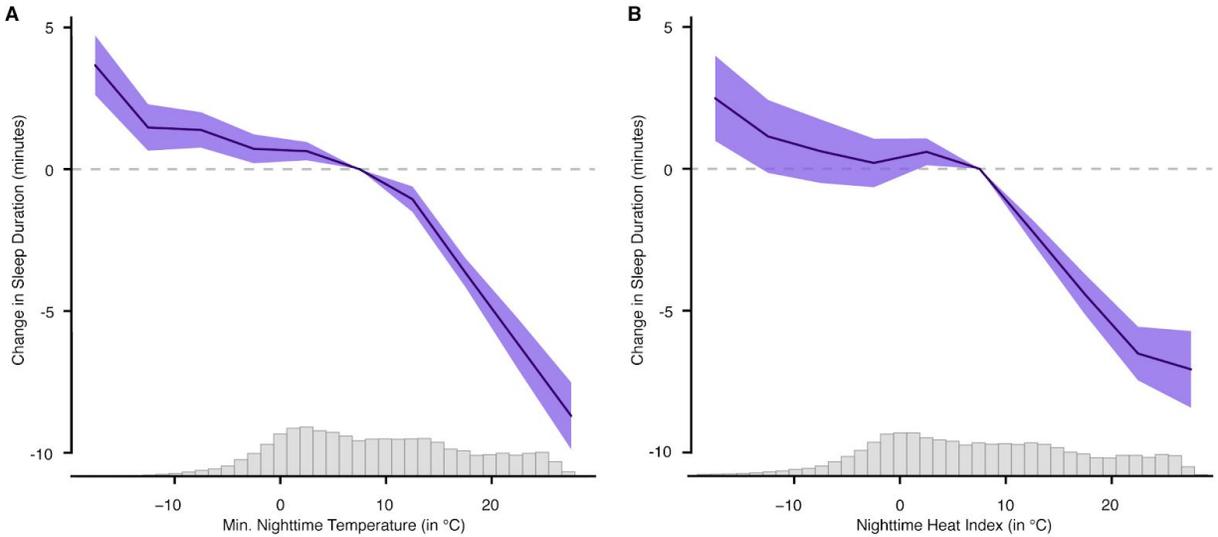

**SI Figure 2** High ambient temperatures and heat index values both reduce sleep duration. **(A)**, Employing gridded reanalysis data (NCEP Reanalysis 2) enables the inclusion of additional users and sleep observations in the estimated relationship between minimum temperature and total time asleep at night. Increases in ambient nighttime temperature reduce sleep duration across the entire temperature distribution, with incrementally larger reductions above 10°C (Supplementary Table 6). Shaded regions represent 95% confidence intervals. Histograms plot the distribution of observed temperatures across over 7 million sleep observations. **(B)**, Elevated heat index values above 10°C reduce sleep attainment, while values below -10°C marginally increase sleep duration (Supplementary Table 15).

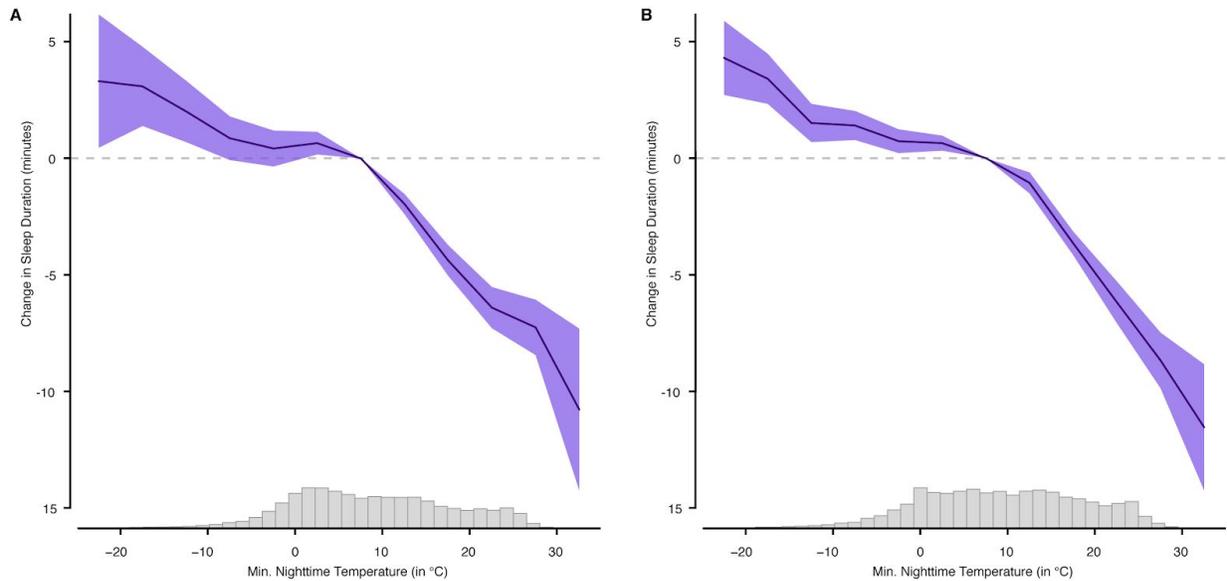

**SI Figure 3** Extremely warm nights generate acute sleep loss. **(A)**, Nighttime temperatures above 30°C (measured with GHCND weather station data) further elevate sleep loss (Supplementary Tables 30, 31). Compared to the reference range of 5 to 10°C, nighttime temperatures above 30°C reduce individual sleep duration by over ten and a half minutes (coefficient: -10.77, $p < 0.001$). Shaded regions represent 95% confidence intervals. Histograms plot the distribution of observed temperatures over the dataset. **(B)**, Estimates generated by substituting GHCND weather station data with globally gridded NCEP Reanalysis 2 meteorological data generate a consistent functional form, but yield larger sleep loss values at higher temperatures. Nighttime temperatures above 30°C decrease individual sleep duration by over eleven and a half minutes compared to temperatures within the 5 - 10°C range (coefficient: -11.54, $p < 0.001$). Conversely, extremely cold nights below -20°C increase sleep duration by over 4 minutes (coefficient: 4.30, $p < 0.01$).

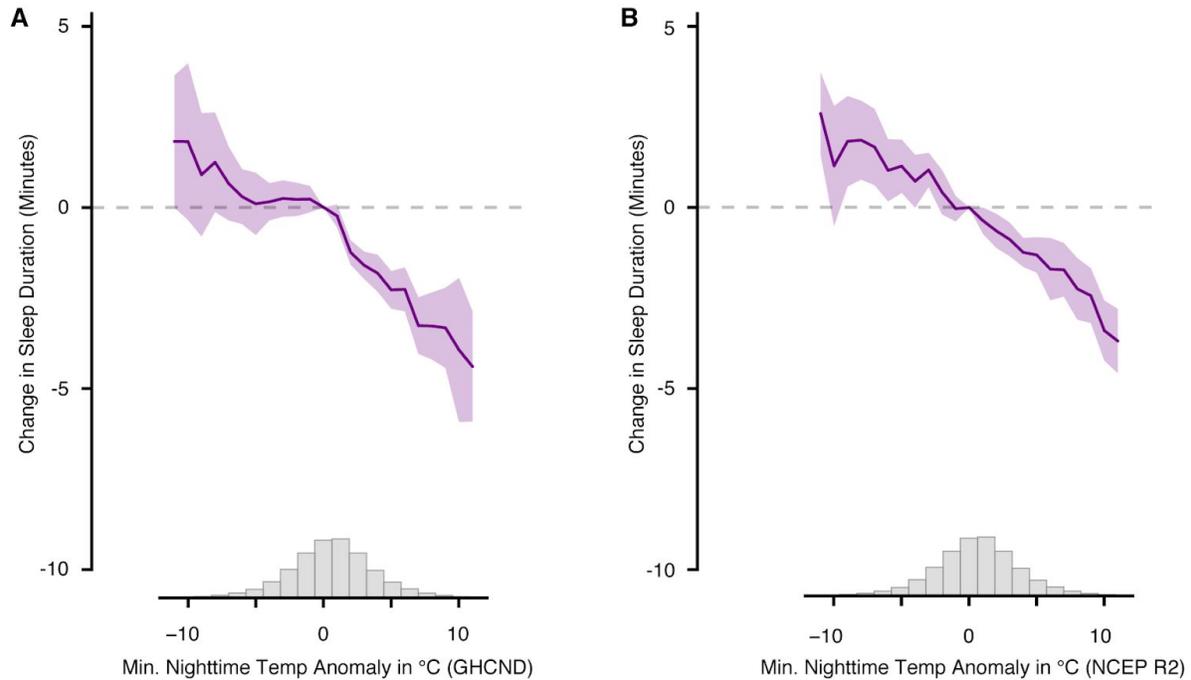

**SI Figure 4** Positive nighttime temperature anomalies reduce sleep duration. **(A)**, Modelling the effect of nighttime temperature increases above local historical averages (1981 - 2010), we find that positive temperature anomalies reduce sleep, consistent with the estimates from our primary model specification (Fig. 2A, Methods Eq. 4a, Supplementary Table 18). Shaded regions represent 95% confidence intervals computed using heteroskedasticity-robust standard errors clustered on the first-level administrative division. Histograms plot the distribution of observed temperature anomalies relative to historical averages. **(B)**, Substituting gridded reanalysis data for station-based temperature measurements produces a consistent functional form with similar estimates. Positive temperature anomalies shorten sleep duration while negative temperature anomalies below local historical averages increase sleep duration.

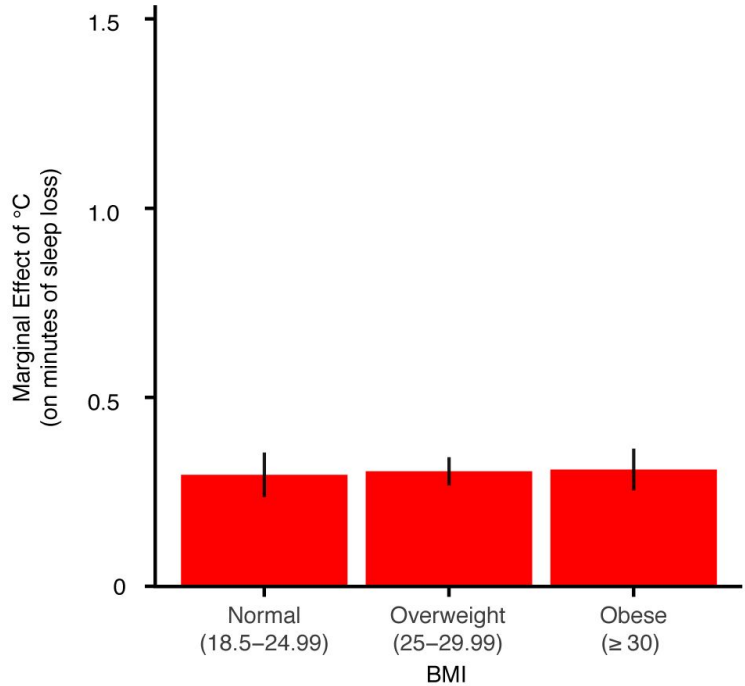

**SI Figure 5** The effect of a 1°C increase in ambient temperature on sleep loss is similar in magnitude across BMI categories. The marginal effects of WHO BMI categories on sleep loss produced by interacting user BMI category with nighttime minimum temperature within our primary model specification (Methods, Eq. 1d). The effect of elevating temperature by 1°C on sleep loss is consistent in magnitude across Normal (n = 2,094,740 observations), Overweight (n = 1,260,247 observations) and Obese (n = 510,742 observations) BMI categories (Supplementary Table 24).

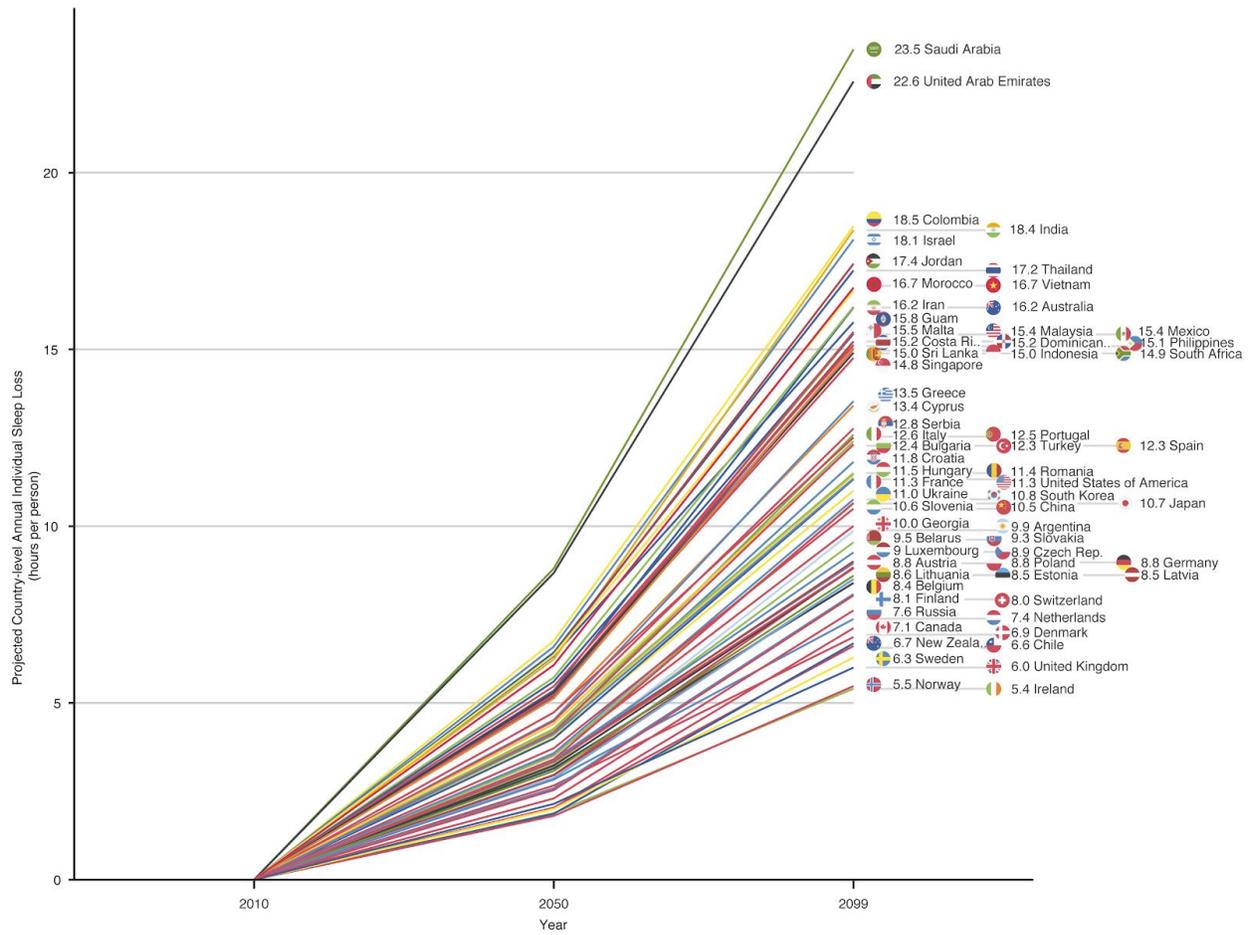

**SI Figure 6** Increases in nighttime temperatures due to climate change are projected to erode human sleep duration across all countries observed, with large regional disparities. Globally averaged projections for the impact of climate change on hours of individual sleep loss, stratified by country. Each colored line depicts the estimated country-level change in per-person sleep loss for the ensemble mean of projected loss across 21 downscaled climate models, averaged across each set of country-level pixels within the dataset, using the fitted values from our spline regression model (Methods, Person-Level Annual Projection Plot). Country icons display national estimates of per-person sleep loss by 2099. Annual sleep loss and country-level disparities increase over time - with the magnitude of relative loss increasing from the middle to the end of the century - due to projected warming under a high emissions scenario (RCP8.5). Without adaptation, countries with relatively warmer climates are expected to sustain greater losses, since estimated marginal sleep loss increases with elevated temperatures (Fig. 2A, SI Fig. 2A).

*Community Health* (2019) doi:10.1007/s10900-019-00731-9.

13. Ford, E. S., Cunningham, T. J. & Croft, J. B. Trends in Self-Reported Sleep Duration among US Adults from 1985 to 2012. *Sleep* **38**, 829–832 (2015).

14. Donat, M. G. *et al.* Updated analyses of temperature and precipitation extreme indices since the beginning of the twentieth century: The HadEX2 dataset. *J. Geophys. Res. Atmospheres* **118**, 2098–2118 (2013).

15. Salamanca, F., Georgescu, M., Mahalov, A., Moustaoui, M. & Wang, M. Anthropogenic heating of the urban environment due to air conditioning. *J. Geophys. Res. Atmospheres* **119**, 5949–5965 (2014).

16. Obradovich, N. & Migliorini, R. Sleep and the human impacts of climate change. *Sleep Med. Rev.* **42**, 1–2 (2018).

17. Rifkin, D. I., Long, M. W. & Perry, M. J. Climate change and sleep: A systematic review of the literature and conceptual framework. *Sleep Med. Rev.* **42**, 3–9 (2018).

18. Schmidt, M. H. The energy allocation function of sleep: A unifying theory of sleep, torpor, and continuous wakefulness. *Neurosci. Biobehav. Rev.* **47**, 122–153 (2014).

19. Harding, E. C., Franks, N. P. & Wisden, W. The Temperature Dependence of Sleep. *Front. Neurosci.* **13**, (2019).

20. Yetish, G. *et al.* Natural Sleep and Its Seasonal Variations in Three Pre-industrial Societies. *Curr. Biol.* **25**, 2862–2868 (2015).

21. Kräuchi, K. The thermophysiological cascade leading to sleep initiation in relation to phase of entrainment. *Sleep Med. Rev.* **11**, 439–451 (2007).

22. Buguet, A. Sleep under extreme environments: Effects of heat and cold exposure, altitude, hyperbaric pressure and microgravity in space. *J. Neurol. Sci.* **262**, 145–152 (2007).

23. Obradovich, N., Migliorini, R., Mednick, S. C. & Fowler, J. H. Nighttime temperature and human
24

**Acknowledgements:** We thank seminar participants at the University of Copenhagen, Copenhagen Center for Social Data Science and Copenhagen Business School Inequality Platform for useful comments.

**Author contributions:** Conceptualization: K.M., A.B.N., S.L., N.O.; Data curation: K.M., A.B.N., S.S.J.; Formal analysis: K.M., N.O.; Funding acquisition: S.L., A.B.N., K.M.; Investigation: K.M., N.O.; Methodology: N.O., K.M.; Software: K.M., N.O., S.S.J., A.B.N.; Visualization: K.M., N.O.; Writing, original draft: K.M.; Writing, review and editing: K.M., N.O., S.L.

**Competing Interests:** The authors declare no competing interests.